\documentclass[12pt,a4paper,pdflatex]{article}
\usepackage{authblk}
\usepackage[T1]{fontenc}
\usepackage[utf8x]{inputenc}
\usepackage{authblk}
\usepackage{amssymb}
\usepackage{esint}
\usepackage[left=2cm,right=2cm,top=3cm,bottom=3cm]{geometry}
\usepackage{graphicx}
\usepackage[figuresright]{rotating}
\usepackage{epsfig}
\usepackage[margin=1cm]{caption}
\def\eref#1{(\ref{#1})}
\usepackage{authblk}
\usepackage{fontenc}
\usepackage{amssymb}
\usepackage{esint}
\usepackage{graphicx}
\usepackage[figuresright]{rotating}
\usepackage{epsfig}
\usepackage{color}

\def\@keyword{}
\newcommand{\keyword}[1]{\gdef\@keyword{#1}}
\def\@pacs{}
\newcommand{\PACS}[1]{\gdef\@pacs{#1}} 
\def\@msc{}
\newcommand{\MSC}[1]{\gdef\@msc{#1}}

\date{}
\begin{document}
\author[1]{M.  Asorey\thanks{asorey@unizar.es}}
\author[2]{L.  Rachwa\l{}\thanks{grzerach@gmail.com}}
\author[3] {I. Shapiro\thanks{ilyashapiro2003@yahoo.com.br}}
\affil[1]{\small Departamento de F\'{\i}sica Te\'orica, 
Universidad de Zaragoza,  E-50009 Zaragoza, Spain.}
\affil[2]{\small Instituto de F\'{\i}sica, Universidade de
Brasilia, C. P. 04455, 70919-970, Bras\'{\i}lia, DF, Brazil.}
\affil[3]{\small  Departamento de F\'{\i}sica--ICE, Universidade Federal de Juiz de Fora, MG, Brazil. }
\title{Unitarity issues in higher derivative field theories}

\maketitle

\abstract
{We analyze the unitarity properties of higher derivative quantum
field theories which are free of ghosts and ultraviolet singularities.
We point out that in spite of the absence of ghosts most of these
theories are not unitary. This result confirms the difficulties of
finding a consistent quantum field theory of quantum gravity.}

\keyword{Quantum gravity, higher derivatives, non-local theories, unitarity, consistency conditions,
Osterwalder-Schrader positivity, Kallen-Lehmann spectral representation.}

\MSC{81T05  	
83D05}

\section{Introduction}

\def\com{\color{magenta}}
\def\cor{\color{black}}
\def\cob{\color{blue}}
\def\con{\color{black}} 
\def\cog{\color{green}}
\def\eref#1{(\ref{#1})}

\hspace{0.5cm} Higher derivative interaction terms arise in the
effective theories of most of local quantum field theories.
However, from a fundamental viewpoint, they are usually avoided
because of their undesirable behaviour concerning causality and
unitarity principles. Fortunately, the Standard Model is a
renormalizable theory with only first and second order derivative
terms in the action, which is consistent with locality, causality
and unitarity requirements. On the other hand, the quantum theory of
Einstein's general relativity is non-renormalizable. A quantum
theory of gravitation requires incorporating terms with a higher
number of derivatives. Dealing with renormalizable or even
UV-finite theories of quantum gravity requires, thus, to address
the possible implications of higher derivative couplings on
fundamental properties of the theory like unitarity and causality.
Local field theory models of quantum gravity involving higher
order derivatives have been classified as generalizations
of Horndeski models \cite{Horndeski}. It is remarkable that the early model of inflation introduced by Starobinsky \cite{Starobinsky}
is embedded in those families. The $R^2$ term, together with the
cosmological constant term $\Lambda$, and the Einstein term $R$
seem to account for the effective behaviour of gravitation at all
scales, from the microworld to cosmology. The fact that all these
terms are local seems to point out a general principle of the
nature of fundamental interactions.

The earliest attempt to introduce a
renormalizable theory of gravity was carried out by Stelle in the
seventies \cite{Stelle}. Super-renormalizable theories of quantum
gravity with higher derivatives were first analyzed two decades
later \cite{alsh, Tomboulis}.

However, the appearance of higher derivatives in local field theories
raises several consistency problems. First, their energy density is not
bounded from below at least for non-degenerate theories because of Ostrogradski
theorem \cite{Ostrogradski}. In the quantum theory, negative energy
states can be traded by negative norm states (or ghosts) leading to
non-unitary theories \cite{GLT,Frolov}.

The standard Pauli-Villars regularization can be formulated as a
higher derivative theory that obviously contains ghosts, which spoil
unitarity and causality properties of the theory. It has been also
shown that any theory with higher order polynomial terms of
D'Alambertian operator with real roots always contains ghosts by
topological reasons \cite{alsh}.

But there are other attempts to overcome the ghost problem by
tricks that remove the negative effects of the ghosts in particular
theories. For instance, Lee-Wick theories skip the appearance
of unphysical ghosts by using polynomials with
complex conjugate poles \cite{leewick,leewick2}
to provide a unitary $S$-matrix of gravitational excitations \cite{LMISh}.

Another way of avoiding the ghost problem is to consider non-local
theories with an infinite number of terms with an arbitrary number of
derivatives, which sum up to form an entire function of the D'Alambertian
operator. Since these functions have no poles on the whole complex
plane, the problem is apparently solved \cite{Tomboulis}.

One has to remember that a consistent field theory should match a
set of basic principles, including relativistic invariance, causality, and unitarity.
These principles impose strong conditions on physical
Green functions, which guarantee that the analytic continuation in
complex time of Wightman functions gives rise to Schwinger
functions in Euclidean time. Thus, a consistent field theory must
define not only regular Wightman functions but also regular
Schwinger functions. Moreover, the Schwinger functions must
satisfy the reflection positivity requirement. In particular the
2-point function must admit a K\"all\'en-Lehmann representation
\cite{Kallen,Lehmann}.

In this paper, we analyze the consistency issue for a large family of
higher derivative theories which are super-renormalizable and ghost-free.
We show that, in spite of their better ultraviolet behaviour
\cite{Tomboulis,Shapiro:2015uxa}
the higher derivative theories of this class fail to match the
complete set of basic requirements and therefore may lead to pathological
quantum field theories. So, even if the non-local ghost-free theories have
interesting quantum and cosmological applications (see e.g. \cite{kmrs} and references therein)
they cannot be considered as candidates for consistent fundamental
theories of quantum gravity.

\section{Non-local finite scalar theories}

Let us consider the simple case of  a scalar  field  theory
\begin{eqnarray}
 S(\phi)
 &=&
 {\textstyle \frac12}  \!\int\!
 \left( (\partial^\mu\phi)^\dagger(\partial_\mu \phi) - m^2|\phi|^2
 - {\textstyle \frac{\lambda}{12}}
 |\phi|^4\right).\label{exp}
\end{eqnarray}

Although it is unclear  the renormalizability of the theory from a
non-perturbative viewpoint,  in four-dimensional spacetime the
theory is perturbatively renormalizable. In perturbation theory
the renormalization prescription requires a regularization of
quadratically and logarithmically divergent Green functions. One
way of regularizing the theory is by introducing an ultraviolet (UV)
regulator. A particular choice is
\begin{eqnarray}
 S(\phi)&=&  {\textstyle \frac12} \! \int \! \left( \phi^\dagger\,e^{({\Box}/\Lambda^2)^s}\left(-\Box- m^2 \right)\phi
 - {\textstyle \frac{\lambda}{12}}
 |\phi|^4\right),
\end{eqnarray}
where $\Box=\partial^\mu\partial_\mu$ is the D'Alambertian operator,
$\Lambda$ is an UV-regulator and the exponent $s$ is a positive number.
The theory is finite in the Euclidean sector because the propagator
 \begin{eqnarray}
\Delta(p)= \frac{e^{-p^{2s}/\Lambda^{2s}}}{p^2+m^2}
\end{eqnarray}
is strongly suppressed in the UV.

However, not all  theories with higher derivatives can be considered as
fundamental theories. The first reason is due to the classical instability
of non-degenerate higher derivative theories \cite{Frolov}, pointed out by the
 Ostrogradski theorem \cite{Ostrogradski}. However,
in the non-interacting case $\lambda=0$, due to  the quadratic
dependence on the fields and  the exponential nature of the higher
derivative operators the energy density is always positive when
evaluated on  solutions of the classical equation of motion, rendering
the free classical theory stable. This property can also be understood
by the fact that one can re-absorb the exponential regulating factors
in a field redefinition
$$
\phi\to \phi'= e^{({\Box}/\Lambda^2)^{{s}/{2}}}\phi,
$$
giving rise to a standard free field theory without any instability.

From a  quantum field theory viewpoint there are some basic principles
that theories must satisfy. In particular
\cite{Glimm-Jaffe, Asorey, Weinbergvol2}:

\begin{itemize}
\item{i)} { \bf Analyticity:} The  $n$-point Wightman functions
$W_n(x_1,x_2,\dots,x_n)=$\hfill\break
$\langle 0|\phi(x_1)\phi(x_2)
\dots\phi(x_n)|0\rangle$
should define
regular distributions  and must admit an analytic continuation to
regular $n$-point Schwinger  functions  $S_n(x_1,x_2,\dots,x_n)
=W_n(\widetilde{x}_1,\widetilde{x}_2,\dots,\widetilde{x}_n)$ in the
Euclidean spacetime coordinates $\widetilde{x}_j=(i  x_j^0, \mathbf{x}_j)$.

\item{ii)} {\bf Reflection Positivity:} The Schwinger functions $S_n$
should satisfy Osterwalder-Schrader reflection positivity property
\cite{OS1,OS2}. In the case of the Euclidean 2-point function this
property implies that
 \begin{equation}
 \int\! \theta f (x){S_2}(x,y)f(y)\geqslant 0
 \label{rp}
  \end{equation}
for any  function $f\in C_0(\mathbb{R}^4_+)$, where $\theta$ is the
time-reversal transform  defined by $\theta f(x^0,\mathbf{x})=
 f^\ast(-x^0,\mathbf{x})$. The functions $f$ in $ C_0(\mathbb{R}_+^4)$
 are continuous functions with compact support in the positive time
  half-space $(x^0>0)$ of $\mathbb{R}^4$.

\item{iii)} {\bf K\"all\'en-Lehmann representation:} The Fourier
transform of the  2-point   \break  Schwinger function should admit a
K\"all\'en-Lehmann representation \cite{Lehmann}
 $$
 \widehat{S}_2(p)=\!\int_0^\infty \! d\mu {\rho(\mu)\over p^2+\mu^2}
 $$
 with a non-negative spectral density $\rho(\mu)\geqslant 0$.
\end{itemize}
{{}

The fundamental principles  of the quantum theory  (i-iii)  together with some other technical requirements 
allow to reconstruct the  quantum Hilbert space of the field theory from  space-time multi-functionals 
$f(x_1,x_2, .... , x_n)$ with support on positive times $t_i>0,\, i=1,2,\dots n$  (Refs. [16,17]). The norm of the 
quantum states is defined by contraction of their indices with those of Schwinger functions (Refs. [16,17]).
If the two-point function is not reflection positive \eref{rp} the induced norm would be negative for  states of the form
$\phi(f)|0\rangle=\int \!dx\, \phi(x) f(x) \,|0\rangle$ and the 
construction of the quantum theory from the Euclidean Schwinger functions would fail. 
In this sense, the failure of reflection positivity implies a
failure of unitarity in the quantum field theory.
}

 In the simplest case $s=1$, the tree-level two-point Schwinger
function
\begin{equation}
S_2(x,y)=\! \int \! \frac{d^4p}{(2\pi)^4}\Delta(p)e^{ip(y-x)}
=\!\int\! \frac{d^4p}{(2\pi)^4}
\frac{e^{-p^{2}/\Lambda^{2}}}{p^2+m^2}  e^{ip(y-x)}
\nonumber
\end{equation}
is finite. However, there is no analytic continuation to the real-time
Minkowski space, as it is the case in the standard theory. The obstruction
is due to the fact that the analytic continuation by Wick rotation requires
the vanishing of the contribution of the contour integrals of the first and
third quadrants at infinite distance, which is not satisfied in this case due
to the existence of essential singularities at the infinity of the complex
energy plane.

The only divergence of the theory arises in the vacuum energy density
\begin{equation}
\Delta \mathcal{E}_0= -\!\int\! \frac{d^4p}{(2\pi)^4}
\frac{\left(p^{0}\right)^2}{\Lambda^2}.\,
\end{equation}
{{}

However, the consistency of the quantum theory requires satisfying the
reflection positivity property \eref{rp}. This means that in particular
\begin{equation}
S_2(\theta x,x)=S_2(-\tau, \mathbf{x};\tau,\mathbf{x})\geqslant 0,
\label{poss}
\end{equation}
which is satisfied by the theory in the case $s=1$, where
\begin{equation}
S_2(\theta x,x)=\!\int\! \frac{d^4p}{(2\pi)^4}
\frac{e^{-p^{2}/\Lambda^{2}}}{p^2+m^2} e^{2 ip^0 \tau} \geqslant 0.
\label{pos}
\end{equation}
However, for theories with higher power exponents $s>1$, this test of
reflection positivity is not matched and the corresponding theories are
not consistent. The breaking of this test of reflection positivity can be explicitly shown by numerical analysis (see Fig.1.)
and rigorously proved by analytic arguments \cite{Widder2, Odlyzko}.
When $0< s\leqslant1$ the exponential term $e^{-p^{2s}/\Lambda^{2s}}$ can be expanded as a non-negative
convex combination of Gaussian terms
\begin{equation}
e^{-p^{2 s}/\Lambda^{2s}}= \int_0^{\infty}\!e^{-x p^{2}/\Lambda^{2}} \rho(x) dx.
\label{bernstein}
\end{equation}
with $\rho(x)\geqslant 0$ for $x>0$ due to Bernstein theorem \cite{Widder2}. Thus,
the positivity of $S_2(\theta x,x)$ follows from that of the Gaussian with $s=1$ (\ref{pos}).
However, for $s>1$ the exponential term does not admit such a representation and the reflection positivity
condition (\ref{poss}) fails (see e.g. Ref. \cite{Odlyzko}).
}

\begin{figure}[tp]
\begin{center}
{\includegraphics[angle=0,width=7.5cm]{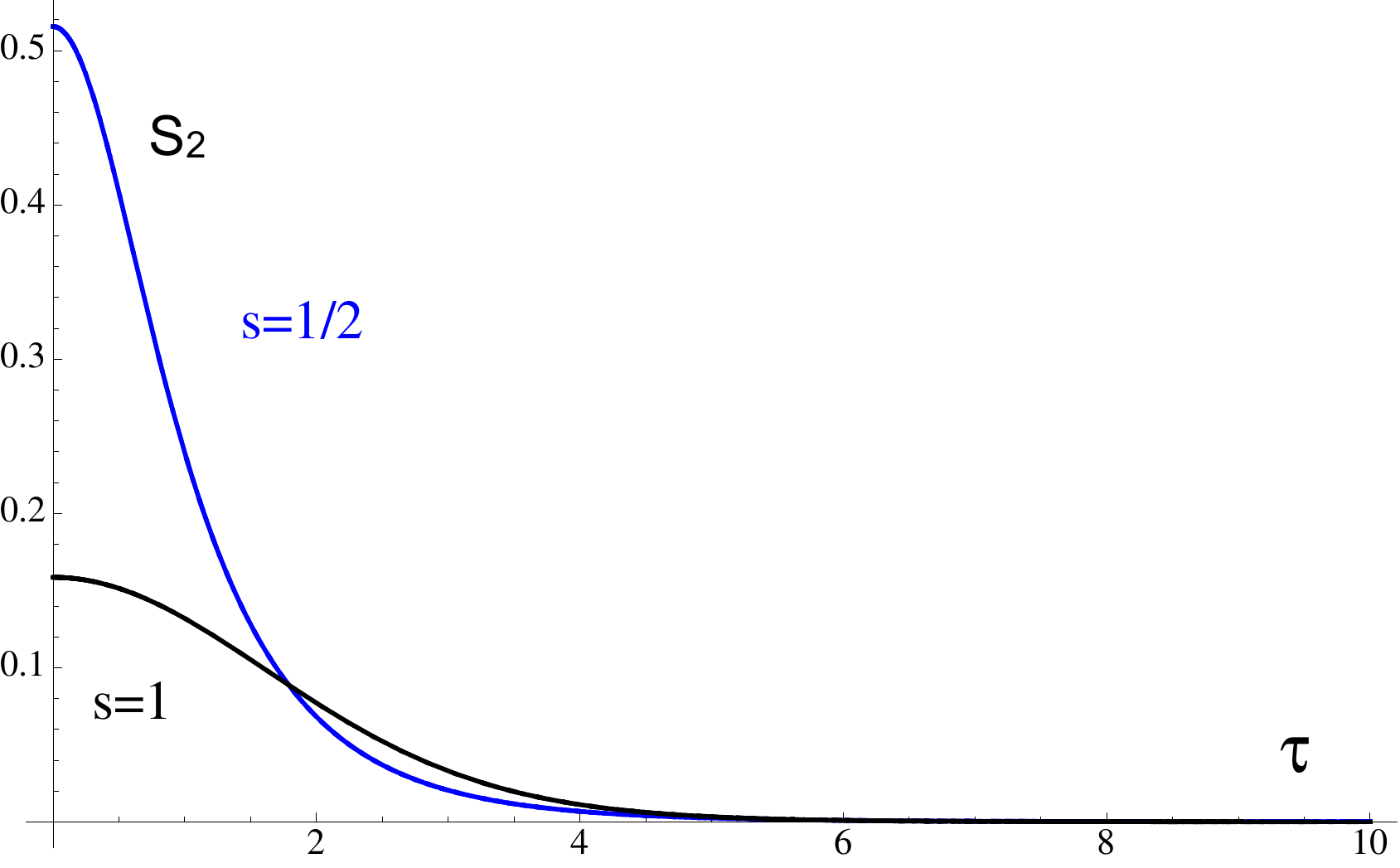}}{\includegraphics[angle=0,width=7.5cm]{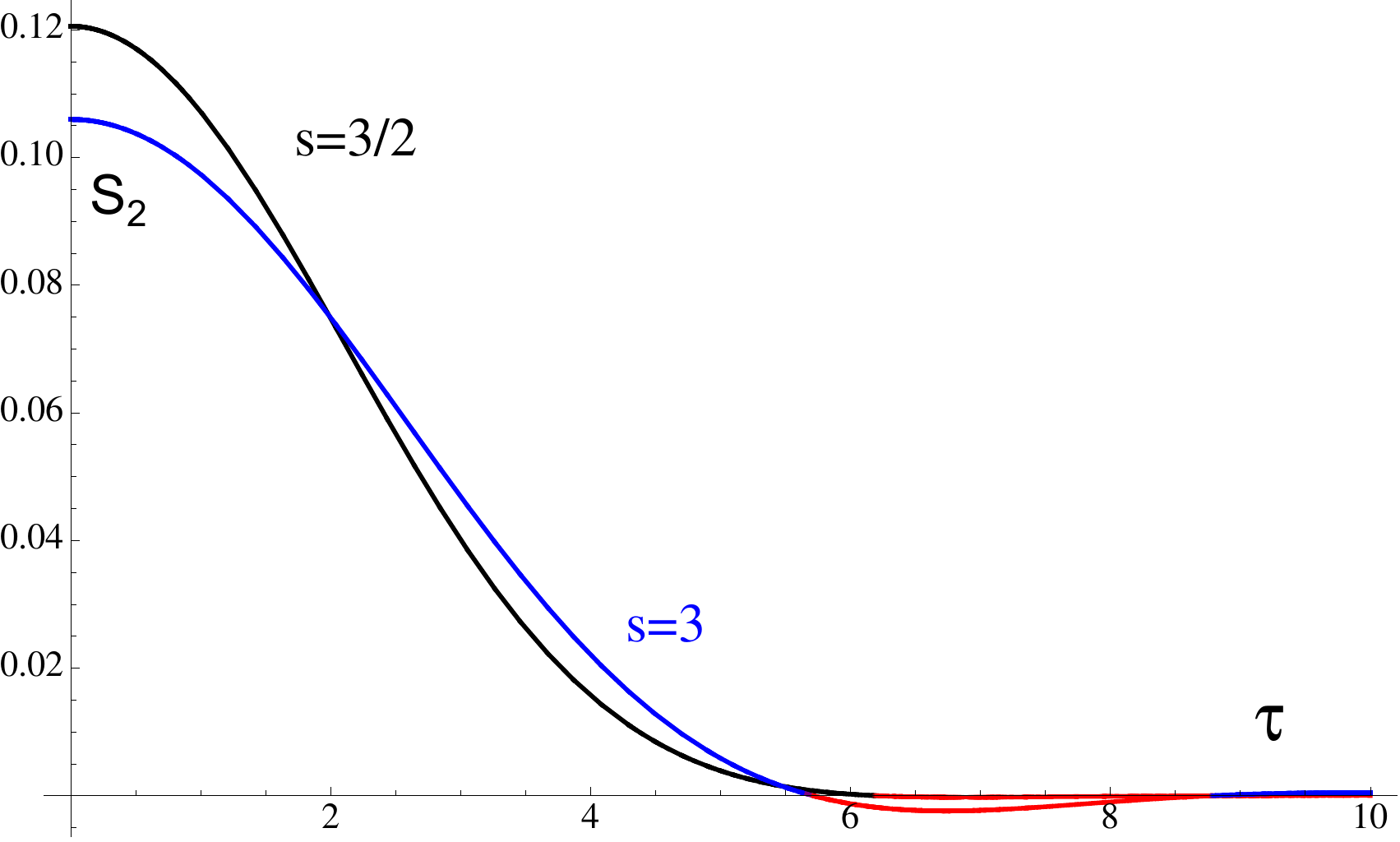}}
\caption{ Time dependence of the (time-reflected) 2-point Schwinger
functions $S_2(-\tau, \mathbf{x};\tau,\mathbf{x})$ for a non-local
scalar field theory with higher derivatives given by \eref{exp}. The red
color indicates the domain where negative values are attained. For values of $s>1$
the function $S_2$ always possesses values of time for which the
function becomes negative. However for $s\leqslant1$ it is always
strictly positive.}
\label{default}
\end{center}
\end{figure}
Thus, even if the theory has no ghosts, it does not mean that it is unitary, as we have shown for $s>1$.

Moreover, even in the case $s=1$ the fact that the 2-point Schwinger
function is strictly positive $S_2(\theta x,x)\geqslant 0$ does not imply
that the corresponding operator \cor $S_2$ is positive definite. \con {{} It simply means that the
diagonal elements of the corresponding matrix are positive. But the
matrix itself might be not positive definite}. In other terms, it simply signifies that
the two-point function is positive for the set of functions $f$ which are highly localized in points
of spacetime. In fact, it can be shown that the operator \cor $S_2$ \con is generally not a
positive definite. It is enough to check that for some
$f\in C_0(\mathbb{R}^4_+)$
\begin{equation}
\int\! \theta f (x){S_2}(x,y)f(y)< 0\,.
\label{rpf}
\end{equation}
This negative result can be derived from the fact that the 2-point
Schwinger function \cor $S_2(x,y)$ \con does not admit a K\"all\'en-Lehmann
representation for any value of $s$. This fact is a consequence of a
mathematical theorem \cite{Roberts} proved by Widder \cite{Widder}
(see also \cite{Krasnikov,Efimov}).

\section{Non-local  
gravitational theories}

The analogous
super-renormalizable  gravitational theories are described by the action
\begin{eqnarray}
S(g)
&=&
\lambda \kappa \!\! \int \!\!  \sqrt{g} +\kappa\!\! \int\!\! \sqrt{g} R
+\alpha \kappa\int\!\!  \sqrt{g}\,  R_{\mu\nu\alpha\beta}
\,e^{({\Box}/\Lambda^2)^s} R^{\mu\nu\alpha\beta}
\nonumber
\\
&+&
\beta\kappa\!\! \int\!\!  \sqrt{g}\,  R_{\mu\nu}
\,e^{({\Box}/\Lambda^2)^s} R^{\mu\nu}
+ \gamma\kappa\!\! \int\!\!  \sqrt{g}\, R
\,e^{({\Box}/\Lambda^2)^s}R\,. 
\label{gravv}
\end{eqnarray}
Classical Ostrogradski instabilities also arise in  these theories
\cite{Frolov}. In perturbation theory, $S(g)$ describes a  theory
of  two self-interacting   massless particles with helicities 2 and 0.
The absence of ghosts can be shown from the absence of unphysical poles in
the 2-point Schwinger function of the perturbation of the gravitational
field:
\begin{eqnarray}
S^{(2)}_{\mu\nu\rho\sigma}(p^2)
&=&
\frac{\mathcal{P}^{(2)}_{\mu\nu\rho\sigma}(p^2)}{p^2- \beta/2\,
e^{p^{2s}/\Lambda^{2s}}p^4-2\,\gamma\, e^{p^{2s}/\Lambda^{2s}} p^4}
\nonumber
\\
&-&
\frac{\mathcal{P}^{(0)}_{\mu\nu\rho\sigma}(p^2)}{p^2+({\beta}/{2}+6\alpha)
\, e^{p^{2s}/\Lambda^{2s}}p^4+8\,\gamma\, e^{p^{2s}/\Lambda^{2s}} p^4}\,,
 \label{grav}
\end{eqnarray}
 where
\begin{eqnarray}  \mathcal{P}^{(2)}_{\mu\nu\rho\sigma}(p^2)
&=&
\textstyle{\frac12} \Big(
\delta_{\mu\rho}\delta_{\nu\sigma}+\delta_{\nu\rho}\delta_{\mu\sigma}\Big)
-\textstyle{\frac13} \delta_{\mu\nu}\delta_{\rho\sigma}
\nonumber
\\
&-&\textstyle{\frac{p_\mu p_\rho}{p^2}\delta_{\nu\sigma}}-{\frac{p_\nu p_\sigma}{p^2}\delta_{\mu\rho}}+{\frac13 } {\frac{p_\mu p_\nu}{p^2}\delta_{\sigma\rho}}\nonumber\\
&-&\textstyle {\frac{p_\mu p_\sigma}{p^2}\delta_{\nu\rho}}-{\frac{p_\nu p_\rho}{p^2}\delta_{\mu\sigma}}+{\frac13 }
{\frac{p_\sigma p_\rho}{p^2}\delta_{\mu\nu}}\nonumber\\
&+&\textstyle
{\frac23 } {\frac{p_\sigma p_\rho p_{\mu}p_{\nu}}{p^4}}\,,
\end{eqnarray}
and
\begin{equation}
\mathcal{P}^{(0)}_{\mu\nu\rho\sigma}(p^2)=\textstyle{\frac13}\left( \delta_{\mu\nu}\delta_{\rho\sigma}
- {\frac{p_\mu p_\nu}{p^2}\delta_{\sigma\rho}}-
{\frac{p_\sigma p_\rho}{p^2}\delta_{\mu\nu}}+
 {\frac{p_\sigma p_\rho p_{\mu}p_{\nu}}{p^4}}\right)
\end{equation}
are the projectors onto the spin-2 and spin-0 ({\it stress scalar}) components of the
2-point function.

What happens in the higher derivative theory is that
the 2-point Schwinger function \eref{grav} does not satisfy the reflection
positivity
$$
\theta S^{(2)}_{\mu\nu\sigma\rho}\geqslant 0.
$$
As we have shown in the scalar case it is enough to prove that
\begin{equation}
S^{(2)}_{ijij}(\theta x,x)
\end{equation}
is not positive for some values of $x\in {\mathbb R}^4$ and $i\neq j$. Indeed,
\begin{eqnarray}
S^{(2)}_{ijij}(\theta x,x)\!\!\!&=& \!\!\!\frac23
\!\int\! \frac{d^4p}{(2\pi)^4} \frac{(1+ p_i^2p_j^2)e^{2 ip^0 \tau}}{p^2- \beta/2\, e^{p^{2s}/\Lambda^{2s}}p^4-2\,\gamma\, e^{p^{2s}/\Lambda^{2s}} p^4}\nonumber\\
&-&\!\!\!  \frac13
\!\int\! \frac{d^4p}{(2\pi)^4} \frac{(1+ p_i^2p_j^2)e^{2 ip^0 \cor\tau\con}}{p^2+({\beta}/{2}+6\alpha) \, e^{p^{2s}/\Lambda^{2s}}p^4+8\,\gamma\, e^{p^{2s}/\Lambda^{2s}} p^4}
\end{eqnarray}
is not always positive definite for the same reasons that \eref{pos} was not positive in the case of scalar fields.
Thus, the absence of ghosts does not guarantee the unitarity of a theory.

{{}
The higher derivatives theories \eref{gravv} are relativistically covariant. If, instead of the exponential form-factor
$e^{({\Box}/\Lambda^2)^s} $, we consider a form-factor $e^{({\Delta}/\Lambda^2)^s} $ with only high spatial derivatives
and only two time derivatives the field theory will be UV-finite and will satisfy the reflection positivity condition \cite{am}. However, these
theories are not relativistically invariant. Thus, it is the combination of Lorentz covariance with higher time derivatives which generates the unitarity issue.
}

\section{Conclusions}

\hspace{0.5cm} The search for a consistent theory of quantum gravity
in field theory framework is still a very challenging open problem. The
promising approach based on higher derivative theories has been always
very attractive because of its nice ultraviolet behaviour and cosmological
applications, including Starobinsky model of inflation.
From a classical field theory viewpoint, many of these models were
discarded by the presence of Ostrogradski instabilities. Quantization
imposes even more constraints because of the appearance of ghost
modes, which introduce violations of unitarity and/or causality. One way
of overcoming these problems is by introducing form-factors with an
analytic dependence on the propagating momenta. This approach is
non-local but has the advantage that enables one to avoid ghosts
with the hope that this will preserve all fundamental properties of a
quantum field theory.

We have shown that for all non-local theories involving an
exponentially decreasing form-factors unitarity is not preserved.
From the Euclidean approach to the quantum theory, we have demonstrated
the existence of states with a negative norm. In summary, the family of
non-local theories of this type cannot be considered as a consistent
approach to quantum gravity. The analysis can be extended to other
higher derivative models of gravity, which is left for a future work.

There are alternatives to the models discussed here. The Lee-Wick
theories are very special cases, which require a more detailed analysis.
A family of models which pass our unitarity check is the
Ho\v{r}ava-Lifshitz family. However, in that case the recovery of
general covariance and Lorentz symmetry in the IR regime is not
guaranteed.

In any case, the problem is very challenging and the restrictions which
our results impose on viable models of quantum gravity constrain very
much the search for a consistent theory. This is not a bad news. It only
means that if there is a quantum theory of gravity based on standard
field theory it would be very special. The search of such a special theory
requires a good understanding of restrictions imposed by the quantum
theory principles.

\section{Acknowledgements}
The work of M.A. is partially supported by Spanish MINECO/FEDER
grant FPA2015-65745-P and DGA-FSE grant 2015-E24/2, and COST
action program QSPACE-MP1405.
The work of L.R. was partially supported by the COST grant for
Short Term Scientific Mission (STSM) within the action
MP1405-37241 for the group of QSPACE program. L.R.
also thanks Instituto de
F\'{\i}sica, Universidade de Bras\'{\i}lia and CAPES for support and warm
hospitality during the final stage of this work.
I. Sh. was partially supported by CNPq and FAPEMIG.


\begin{thebibliography}{99}

\bibitem{Horndeski}
G. W. Horndeski, Int. J. Theor. Phys. {\bf 10} (1974) 363-384
\bibitem{Starobinsky}
A.A. Starobinsky, 
Phys. Lett. B91 (1980) 99-102.

\bibitem{Stelle}
K.S. Stelle, 
Gen. Rel. Grav. 9 (1978) 353-371

\bibitem{alsh}
M.  Asorey, J. L. Lopez, and I. L. Shapiro, Int. J. Mod. Phys. {\bf A 12} (1997) 5711-5734.

\bibitem{Tomboulis} E. T. Tomboulis,
Superrenormalisable gauge and gravitational theories,
Arxiv preprint:  hep-th/9702146 (1997).

\bibitem{Ostrogradski}
M.V. Ostrogradski, 
Mem. Acad. St. Petersburg 6 (1850) 385-517.

\bibitem{GLT} D. M. Gitman and I.V. Tyutin,
{\it Quantization of Fields with Constraints,}
Springer (1990).

\bibitem{Frolov}  V. P. Frolov and A. Zelnikov,	Phys. Rev. D 93, 105048 (2016).

\bibitem{leewick} T. D. Lee and G. C. Wick, 
Nucl. Phys. {\bf B 9} (1969) 209-243.
\bibitem{leewick2} T. D. Lee and G. C. Wick, 
Phys. Rev. {\bf D2} (1970) 1033-1048.

\bibitem{LMISh}
  L.~Modesto and I.L.~Shapiro,
Phys. Lett. B {\bf 755} (2016) 279.

\bibitem{Kallen}
G. K\"allen, Helv. Phys. Acta {\bf 25}(1952) 417-434.

\bibitem{Lehmann}
H. Lehmann, Nuovo Cimento 11 (1954) 342-357.

\bibitem{Shapiro:2015uxa} I.L.~Shapiro,
Phys. Lett. B {\bf 744} (2015) 67.

\bibitem{kmrs}
A.S. Koshelev, L. Modesto, L. Rachwal, and A.A. Starobinsky,
JHEP {\bf 1611} (2016) 067.

\bibitem{Glimm-Jaffe} J. Glimm and A. Jaffe,
{\sl Quantum Physics}, Springer-Verlag (1981).

 \bibitem{Asorey}M. Asorey, {\sl A concise introduction to
 Quantum Field Theory}, in press (2017).

\bibitem{Weinbergvol2}  S. Weinberg, The Quantum Theory of Fields, vol. 2, Cambridge Univ. Press (1995).

\bibitem{OS1}
K. Osterwalder and R. Schrader, Commun. Math. Phys. 31 (1973) 83.
\bibitem{OS2}
 K. Osterwalder and R. Schrader, Commun. Math. Phys. 42 (1975) 281.

{{}
\bibitem{Widder2}  D.V. Widder,The Laplace Transform. Princeton:Princeton University Press 1941.

\bibitem{Odlyzko} N. D. Elkies, A. M. Odlyzko, and J. A. Rush, Invent. Math. 105 (1991) 613.
}

\bibitem{Roberts} C. D. Roberts, Prog. Part. Nucl. Phys. 61 (2008) 50.

\bibitem{Widder}D. V. Widder, Bull. Amer. Math. Soc. 40 (1934) 321.


\bibitem{Krasnikov}
N. V. Krasnikov, 
Theor. Math. Phys. {\bf  73} (1987) 1184.
\bibitem{Efimov}
G. V.  Efimov, Commun. Math. Phys. {\bf 5} (1967) 42-56.

{{}

\bibitem{am}M.  Asorey and P. K. Mitter, 
Commun. Math. Phys.  {\bf 80} (1981) 43-58.
}

	
\end{thebibliography}
\end{document}